\documentclass[amsmath,amssymb,preprint]{revtex4}   
\begin{document}    
\title{{Vafa-Witten theorem and Lee-Yang singularities}}   
\author{\bf {M. Aguado$^a$ and M. Asorey$^{b}$}}   
\address{$^{a)}$Max-Planck-Institut f\"ur  Quantenoptik. 
Hans-Kopfermann-Str. 1. D-85748 Garching, Germany}
\address{$^{b)}$Departamento de
F\'{\i}sica Te\'orica. Facultad de Ciencias  
Universidad de Zaragoza.  50009 Zaragoza. Spain
}   

\newcommand\traza{\mathrm{Tr} \,}
\renewcommand\Re{\,\mathrm{Re}\,}
\renewcommand\Im{\,\mathrm{Im}\,}
\def\CAG{{\cal A/\cal G}} \def\CO{{\cal O}} 
\def\CA{{\cal A}} \def\CC{{\cal C}} \def\CF{{\cal F}} \def\CG{{\cal    
G}} \def\CL{{\cal L}} \def\CH{{\cal H}} \def\CI{{\cal I}}    
\def\CU{{\cal U}} \def\CB{{\cal B}} \def\CR{{\cal R}} \def\CD{{\cal    
D}} \def\CT{{\cal T}} \def\CK{{\cal K}}    
\def\e#1{{\rm e}^{^{\textstyle#1}}}    
\def\grad#1{\,\nabla\!_{{#1}}\,}    
\def\gradgrad#1#2{\,\nabla\!_{{#1}}\nabla\!_{{#2}}\,}    
\def\ph{\varphi}    
\def\semi{;}
\def\db{\bar\partial}           
\def\Dsl{\,\raise.15ex\hbox{/}\mkern-13.5mu D} 
\def\dsl{\raise.15ex\hbox{/}\kern-.57em\partial}    
\def\psibar{\overline\psi}    
\def\om#1#2{\omega^{#1}{}_{#2}}    
\def\vev#1{\langle #1 \rangle}    
\def\lform{\hbox{$\sqcup$}\llap{\hbox{$\sqcap$}}}    
\def\darr#1{\raise1.5ex\hbox{$\leftrightarrow$}\mkern-16.5mu #1}    
\def\lie{\hbox{\it\$}} 
\def\ha{{1\over2}}    
\def\half{{\textstyle{1\over2}}} 
\def\roughly#1{\raise.3ex\hbox{$#1$\kern-.75em\lower1ex\hbox{$\sim$}}}    
\def\inbar{\,\vrule height1.5ex width.4pt depth0pt}    
\def\minbar{\,\vrule height1.0ex width.4pt depth0pt}    
\def\IB{\relax{\rm I\kern-.18em B}}    
\def\IC{\relax\hbox{$\inbar\kern-.3em{\rm C}$}} \def\ID{\relax{\rm    
I\kern-.18em D}}    
\def\IE{\relax{\rm I\kern-.18em E}}    
\def\IF{\relax{\rm I\kern-.18em F}}    
\def\IG{\relax\hbox{$\inbar\kern-.3em{\rm G}$}}    
\def\IH{{\Bbb H}}    
\def\II{\relax{\rm I\kern-.18em I}}    
\def\IK{\relax{\rm I\kern-.18em K}}    
\def\IL{\relax{\rm I\kern-.18em L}}    
\def\IM{\relax{\rm I\kern-.18em M}}    
\def\IN{\relax{\rm I\kern-.18em N}}    
\def\IO{\relax\hbox{$\inbar\kern-.3em{\rm O}$}}    
\def\IP{\relax{\rm I\kern-.18em P}}    
\def\IQ{\relax\hbox{$\inbar\kern-.3em{\rm Q}$}}    
\def\IR{{\Bbb R}}    
\font\cmss=cmss10 \font\cmsss=cmss10 at 10truept
\font\rms=cmmi7 at 8truept
\font\eightpt=cmr8 at 8truept
\font\af=msbm10 \font\afs=msbm8    
\mathchardef\imath="717B    
\def\Z{{\af Z}}    
\def\R{{\af R}}    
\def\T{{\af T}}    
\def\C{{\af C}}    
\def\F{{\af F}}    
\def\IZ{\relax\ifmmode\mathchoice    
{\hbox{\cmss Z\kern-.4em Z}}{\hbox{\cmss Z\kern-.4em Z}}    
{\lower.9pt\hbox{\cmsss Z\kern-.36em Z}}    
{\lower1.2pt\hbox{\cmsss Z\kern-.36em Z}}\else{\cmss Z\kern-.4em    
Z}\fi} \def\ZZ{{\Bbb Z}}    
\def\IGa{\relax\hbox{${\rm I}\kern-.18em\Gamma$}}    
\def\IPi{\relax\hbox{${\rm I}\kern-.18em\Pi$}}    
\def\ITh{\relax\hbox{$\inbar\kern-.3em\Theta$}}    
\def\IOm{\relax\hbox{$\inbar\kern-3.00pt\Omega$}}    
\def\semi{;\hfil\break}    
\def\CA{{\cal A}}\def\CCA{$\CA$}    
\def\CC{{\cal C}}\def\CCC{$\CC$}    
\def\CT{{\cal T}}\def\CCT{$\CT$}    
\def\CQ{{\cal Q}}    
\def\CS{{\cal S}}    
\def\CP{{\cal P}}\def\CCP{$\cal P$}    
\def\CO{{\cal O}}\def\CCO{$\CO$}    
\def\CM{{\cal M}}\def\CCM{$\CM$}    
\def\CMH{\widehat\CM}\def\CCMH{$\CMH$}    
\def\CMB{\overline\CM}\def\CCMB{$\CMB$}    
\def\CH{{\cal H}}\def\CCH{$\CH$}    
\def\CL{{\cal L}}\def\CCL{$\CL$}    
\def\CS{{\cal S}}\def\CCS{$\CS$}    
\def\CX{{\cal X}}    
\def\CE{{\cal E}}\def\CCE{$\CE$}    
\def\CV{{\cal V}}\def\CCV{$\CV$}    
\def\CU{{\cal U}}\def\CCU{$\CU$}    
\def\CF{{\cal F}}\def\CCF{$\CF$}    
\def\CG{{\cal G}}\def\CCG{$\CG$}    
\def\CN{{\cal N}}\def\CCN{$\CN$}    
\def\CD{{\cal D}}\def\CCD{$\CD$}    
\def\CZ{{\cal Z}}\def\CCZ{$\CZ$}    
\def\cte{\rm cte}    
\def\cs{W(z)}    
    
\def\Tr{{\rm Tr}}    
\def\tr{{\rm tr}}    
\def\Nabla{{\Bbb D}}    
\def\sph{{z_{\rm sph}}}    
\def\asph{{\widetilde{z}_{\rm sph}}}    
\def\vac{{z_{\rm vac}}}    
\def\be{\begin{eqnarray}}    
\def\ee{\end{eqnarray}}    
\def\th{$\theta$}    
\def\cp{{\af CP}$^{\rm N}$}    
\def\cpp{{{\rm S}^{\rm 2N-1}}}    
\def\cpa{{\af CP}$^{\rm 1}$} 
\def\sm {sigma model}    
\def\Im {{\, \rm Im\, }}    
\def\Re {{\, \rm Re\, }}      
\vskip .5in   
\normalsize   
\begin{abstract}    
We prove the analyticity  of the finite volume  QCD  partition    
function  for complex values of  the $\theta$-vacuum 
parameter. The absence  of  singularities
different from Lee-Yang zeros only permits $\wedge$ cusp singularities in
the vacuum energy density and never $\vee$ cusps.
This fact together with the Vafa-Witten diamagnetic inequality
implies the vanishing of the density of Lee-Yang zeros at $\vartheta=0$ 
and has an important consequence:
the absence of a first order phase transition at  $\theta=0$.
The result  provides a key 
missing link in the Vafa-Witten proof of parity symmetry conservation 
in vector-like gauge theories and follows from renormalizability, unitarity, 
positivity and existence of BPS bounds. 
Generalizations of this theorem to 
other physical systems are also discussed, with particular interest focused 
on the non-linear \cp\ sigma model.
    
\end{abstract}    
\hyphenation{}    
\maketitle \thispagestyle{empty} 
   
The Vafa-Witten theorem \cite{vw} is one of the very few    
non-perturbative analytic results of QCD. In absence of    
supersymmetry, non-Abelian continuum gauge theories in 3+1    
dimensions do exhibit many very interesting non-perturbative    
phenomena which have been  so far elusive to analytic approaches.    
However, by a very simple and elegant argument Vafa and Witten    
\cite{vw} explained why we do not observe spontaneous  violation    
of parity symmetry, not only in QCD but in any gauge theory with    
Dirac fermions in 3+1 space-time dimensions.
The relevance of the physical implications of this theorem 
  require    to further
clarify some  questions  raised about the Vafa-Witten   
argument  (see \cite{ch} for a review and    
references therein).    
    
Vafa and Witten claimed that their argument should hold for any parity    
violating order parameter, but this might not be true in general. In    
fact, order parameters including higher order composite operators    
might be non-renormalizable and it is unclear how the argument    
proceeds in such a case \cite{ch,wu}. On the other hand, non-local order parameters    
like a Wilson loop around a very small compactified space direction acquire  
non-trivial vev's. Thus, one should at least restrict the type of    
order parameter to  only local operators which are renormalizable,    
parity odd and preserve all other symmetries of the system (including    
gauge symmetry and Lorentz invariance). 
One of the order parameters  matching all these    
requirements in 3+1 dimensions is the topological density    
$\epsilon^{\mu\nu\sigma\rho} F_{\mu\nu} F_{\sigma\rho}$.  However,    
even in that case the argument of Vafa and Witten can be questioned    
\cite{ag,Ji}. The aim of this note is to rigorously prove the    
Vafa-Witten theorem for this particular order parameter.    
    
In this case there is no doubt about the existence of a well defined    
free energy for any real value of the $\theta$ parameter. This is    
guaranteed by unitarity of the theory, which is translated into    
Osterwalder-Schrader (OS) positivity of the Euclidean functional    
measure of the theory with real \th-terms \cite{es,am}. We recall that    
OS positivity is not just the standard positivity used in the    
Vafa-Witten argument. This special property excludes the possibility    
that pathological scenarios where the free energy is ill-defined    
\cite{ag} might occur in these theories. The Vafa-Witten    
argument can be correctly applied to show that the free energy density    
$\CE_\theta=E_\theta/V$  at $\theta=0$ is lower or equal than at 
$\theta\neq 0$, i.e. $\CE_0\leq \CE_\theta$.    
    
However, this fact does not mean that the theorem is already proven.    
One needs to exclude the existence of a cusp in the energy density at    
$\theta=0$. The appearance of the cusp would signal the existence of a    
first order phase transition and spontaneous breaking of parity at    
$\theta=0$ without violation of the Vafa-Witten inequality for free    
energies. The only extra property one needs to prove is smoothness of    
$\CE_\theta$ at $\theta=0$, which would make the    
existence of such a cusp  not possible.  In other terms, $\theta=0$ is    
always an absolute minimum of the free energy density but only if    
$\CE_\theta$ is smooth we will have $\CE'_0=0$. Only in that case we    
can make sure that the vacuum expectation value of the topological    
density vanishes and the Vafa-Witten theorem holds  for    
this particular order parameter.    
   
In fact, due to  Bragg symmetry $\CE_{\pi+\theta}=\CE_{\pi-\theta}$, which   
follows from CP symmetry $\CE_\theta=\CE_{ -\theta}$ and   
$\theta$-periodicity $\CE_\theta=\CE_{\theta+2\pi}$, the two   
classically CP symmetric points $\theta=0,\pi$ are extremals of the   
vacuum energy density. However, $\CE_\theta$ is not always smooth at   
$\theta=\pi$: there are cases where CP is spontaneously broken at that   
point. In principle, the same pathology could appear also at   
$\theta=0$.

In order to prove smoothness, the role of Lee-Yang singularities    
becomes very relevant.  Because of Bragg symmetry    
and unitarity, non-analyticities are harder to trace in    
the pure real $\theta$ sector.  Non-analyticity in $\theta$ can    
be better inferred from the lack of an analytic continuation into    
the complex $\theta$ plane. Indeed, in that case OS positivity (i.e.    
unitarity) is not preserved and there is even no guarantee that    
the free energy is well defined, in which  case it will not make    
sense to talk about smoothness. Essentially, there are two    
possible ways in which the $\theta$ theory might give rise to    
pathological non-analyticities in the complex sector. One    
possibility is that the analytic continuation to complex values    
of \th\ is not defined at all, i.e.  the partition function    
itself becomes divergent.  We will see that this possibility does    
not occur in gauge theories. The other possibility is that the    
partition function exhibits a sequence of zeros converging to    
$\theta=0$ in the infinite volume limit. This is the scenario    
advocated by Lee and Yang \cite{ly} which signals in many cases    
the existence of a phase transition like in the Ising model    
in presence of an external (imaginary) magnetic field. 
    
The partition function splits into a sum of functional integrals over    
the different topological sectors    
\begin{widetext}   
\be    
\CZ_\theta (g)=\sum_{q=-\infty}^\infty    
{\rm e}^{-q\Im\theta +\dot \imath q\Re\theta}    
\int\mkern-18mu{\lower3ex\hbox{${}_{c_2(A)=q}$}}    
\mkern-10mu{ \delta A}\  {\rm e}^{-S_{\rm YM}(q)/2g^2}    
\det (\Dsl+m)    
\label{ptt}    
\ee    
%
\end{widetext}     
\noindent   
In a strict sense, the functional integral is UV divergent but it    
can be regularized in such a way that its positivity properties are    
preserved \cite{aff1}. We will consider such a geometrical    
regularization whenever it is required. On the other hand, in the    
infinite volume limit $\CZ_\theta (g)$ vanishes unless the vacuum    
energy is renormalized to 0. But it is just the dependence on    
$\theta$ of the vacuum energy what we want to analyze. Therefore,    
we will consider throughout the paper a compact space-time with large    
but finite volume $VT<\infty$.    
    
Positivity of the Yang-Mills Euclidean measure \cite{amm,aff1} and    
fermionic determinants \cite{vw} implies the existence of a real    
effective action $S_{\rm eff}(g,q)= S_{\rm YM}(q)/2g^2 -\quad \log\det    
(\Dsl+m)$.  On the other hand the Yang-Mills action $S_{\rm YM}(q)$ is    
bounded below by the BPS bound $S_{\rm YM}(q)\geq 8 \pi^2|q|$ in each    
$q$-topological sector.  We shall see that the total effective action    
$S_{\rm eff}(g,q)$ is also bounded from below by a similar linear    
bound involving the absolute value of the topological charge $|q|$. In    
the regularized theory, the coefficient of this linear bound increases   
with the UV regulating scale.  This growth of the effective action is    
enhanced in the chiral limit of light quark masses by the contribution    
of fermionic determinants.  In general, from positivity of fermionic    
determinants and Yang-Mills measure and the BPS bounds we have the    
inequality    
\begin{widetext}     
\be    
\left|\CZ_\theta (g)\right|\leq    
\sum_{q=-\infty}^\infty    
{\rm e}^{|q\Im\theta| } \int\mkern-18mu{\lower3ex\hbox{${}_{c_2(A)=q}$}}    
\mkern-10mu{ \delta A}\  {\rm e}^{-S_{\rm eff}(g,q)}\leq    
\sum_{q=-\infty}^\infty    
\int\mkern-18mu{\lower3ex\hbox{${}_{c_2(A)=q}$}}    
\mkern-10mu{ \delta A}\  {\rm e}^{-S_{\rm eff}(\tilde{g},q)}=    
\CZ_0 (\tilde{g})    
\label{arth}    
\ee    
%
\end{widetext}     
\noindent   
where $1/\tilde{g}^2=1/g^2-(1/8\pi^2) |\Im\theta|$. This shift of the    
gauge coupling constant can be considered as a change of the    
renormalization scale.  This shows that the partition function of    
the original theory with a complex \th-term is bounded by the    
partition function of a similar theory with \th$=0$ but with a    
different coupling constant. Since the theory at $\theta=0$ is    
unitary and renormalizable, its partition function $\CZ_0$ is    
finite and  from (\ref{arth}) it follows that    
$\left|\CZ_\theta\right|<\infty$ is also finite for small values    
of $\Im\theta$.  If in addition the theory is asymptotically free,    
then the renormalization of $g^2$ in the UV fixed point can absorb    
any value of $\Im\theta$, which implies that the radius of convergence    
of the sum (\ref{ptt}) in the complex $\theta$ plane, in fact, becomes    
infinity. To make the argument more precise one should consider the    
regularized theory. The analyticity of $\CZ_\theta$  also
appears naturally in lattice regularizations, 
even with  non-topological regularizations of the
 $\theta$--term. However,  to prove of absence of poles 
 the  topological charge bound (\ref{arth}) is essential. 
 The bare coupling constant $g$ has to be fine tuned    
according to the renormalization group to yield the appropriate continuum    
limit. But because of asymptotic freedom this coupling goes to    
zero as the UV regulator is removed and the shift from $g$ to    
$\tilde{g}$ induced by $\Im \theta$ simply implies a change in the    
effective scale of the continuum theory. However, the proof of the 
Vafa-Witten theorem only requires a finite radius of convergence of    
$\CZ_\theta$ around $\theta=0$. This feature does    
not require asymptotic freedom, just renormalizability.

We remark that one cannot use the same kind of arguments for other    
order parameters. For instance, the partition function associated to    
higher odd powers of the topological density might diverge for any    
complex value of the perturbation parameter, as they are    
non-renormalizable. A similar remark applies to the perturbation by a    
chemical potential term in theories at finite density. In that case    
the missing property is the lack of a BPS bound for the perturbation.

The only remaining possible source of non-analyticity is the presence
of Lee-Yang singularities, i.e.  zeros in the partition function 
$\CZ_\theta$ which could prevent the existence of a unique limit of 
the free energy $\log \CZ_\theta$ or its derivatives at $\theta=0$.

Because of  Bragg symmetry and $\theta$--periodicity $\CZ_\theta$
is a function of  $\sin^2 ( \vartheta / 2 )$ and its  zeros  
come in 4-plets $\{ \vartheta_n, \, -\vartheta_n, \,
\vartheta_n^\ast, \, -\vartheta_n^\ast \}$ in the modular strip 
$\vartheta \in (-\pi, \, \pi] \times \mathbb{R}$) of the
complex plane.   They can be parametrized by their representatives 
in the positive quadrant $\vartheta_n$ , which cannot lie on the real or imaginary 
axes for any finite volume.

Only zeros converging to the origin in the thermodynamic limit can
give rise to a cusp.  Moreover, only an
infinity of such zeros can produce the cusp.

The Weierstrass  factorization theorem implies that the vacuum energy density
\begin{equation}\label{easy:eq:vacuumenergydef}
 { \cal{E}}_V ( \vartheta )
= 
 - \,
  \frac{ 1 }{ V }
  \ln Z_V ( \vartheta ) ,
\end{equation}
can be rewritten in  the functional form \cite{ww,nw}
%
\begin{widetext}   
\begin{equation}\label{easy:eq:vacuumenergy}
  {\mathcal{E}}_V ( \vartheta )
= 
 - \,
  \frac{ 2 }{ V }
  \Re
  \sum_{ n = 1 }^\infty
  \rho_n
  \left\{
    \ln
      \left(
        1
       - \,
        \frac{
          \sin^2 ( \vartheta / 2 )
        }{
          \sin^2 ( \vartheta_n / 2 )
        }
      \right)
    +
    h_n (\sin^2 ( \vartheta / 2 ), V )
  \right\}+ g_{_V}(\sin^2 ( \vartheta / 2 )) ,
\end{equation}
\end{widetext}  
\noindent
where  $\vartheta_n$ denotes the order $n$ Lee-Yang zero in the positive quadrant, 
 $\rho_n$  its degeneracy
and  $h_n$ and  $g$ are, respectively,
polynomial and analytic functions of $   \sin^2 ( \vartheta / 2 )$. The polynomial
 functions  $h_n$ are necessary to 
compensate the divergent effect  of zeros with slow approach to the
infinite modulus for a given volume.

In the thermodynamic limit the degeneracy factor $\rho_n$  of zeros 
converging to the origin defines a density of zeros $\rho ( z )$  as a function of
the continuous variable $z =  \vartheta_n  / V$.  Then, in the thermodynamic limit
%
\begin{widetext}   
\begin{eqnarray}\label{easy:eq:vacuumenergycontinuum}
 { \mathcal{E}} ( \vartheta )
&=&
 - \,
  2
  \Re
\int_{0}^{\pi} \int_0^\infty   {\mathrm{d}} z_1 \,{\mathrm{d}} {z_2} \,
  \rho ( \sin^2 (  z / 2 ) )
  \left\{
    \ln
      \left(
        1
       - \,
        \frac{
          \sin^2 ( \vartheta / 2 )
        }{
          \sin^2 (  z / 2  )
        }
      \right)   +
    h (    \sin^2 ( \vartheta / 2 ), \, z )
  \right\} \nonumber\\
   &&  +\  g(\sin^2 ( \vartheta / 2 )) .
\end{eqnarray}
\end{widetext}  
  Notice that only if $\rho(0)\neq 0$ the contribution 
  from the logarithm at $z=0$ 
  could give rise to a cusp at $\vartheta=0$.
     The rest of the contributions from the integral 
      have continuous derivatives for real $\vartheta$ at
  $\vartheta = 0$.
  In fact, the logarithmic term in (\ref{easy:eq:vacuumenergycontinuum})
   corresponds to a Coulomb
  potential, in two dimensions, generated by the  charge
  density  $- 2 \rho (z)$. The  cusp at $\theta=0$ is only possible if there is a 
  non-vanishing linear charge density at $\theta=0$  \cite{evans}\cite{evans1},
   i.e.  $\rho(z) \sim \rho_0 \,  \delta(z_1- m z_2^\beta)$ for  $|z|\ll 1$ with 
   $\rho_0\neq 0$. A cusp corresponds to a 
  gap in the  potential derivative, which is proportional  to the charge density 
  $-2\rho_0$.  Since this charge density is non-positive, the corresponding
  potential is always repulsive, which means that  any  cusp  is always a 
  $\wedge$ cusp instead of a $\vee$ cusp.  Since the vacuum
  energy density $ { \mathcal{E}}(\vartheta)$ has a local minimum at $\theta=0$
  because of the Vafa-Witten inequality such a cusp cannot exist. 
  The theorem follows from two very simple facts, the Vafa-Witten inequality and the
  repulsive character of the  negative charge
  density defined by $-2 \rho (z)$, which follows from the fact that the partition function
  $Z_V ( \vartheta) $ is analytic and has zeros but  no poles.  The only way of 
  having a a $\vee$ cusp at $\vartheta=0$ is with an attractive charge density
   which can only  only appear from poles in $Z_V ( \vartheta)$. In fact, this is what
   happens in 1+1 massless scalar field theories with pseudo-periodic boundary conditions
   \cite{adm} where the partition function is non-analytic in the complex plane.
However,  in  QCD the analyticity of the partition function is
 based in very simple fundamental properties of the theory: unitarity,    
renormalizability, positivity and existence of BPS bounds.
This makes  impossible  the existence of a cusp in  the vacuum energy density  
at $\vartheta=0$. In summary, the  theory cannot undergo a first order phase transition
at $\theta=0$ with parity symmetry breaking  and we have
a complete proof of the Vafa-Witten theorem. 

From these arguments we cannot, however, exclude the existence of first order 
phase transitions for other values of $\theta\neq 0$.  In    
particular, we cannot discard  CP symmetry breaking for
$\theta=\pi$.

We remark that many other theories share the same properties and    
therefore the Vafa-Witten theorem can also be extended for all of    
them. One particularly interesting case is that of \cp\ non-linear    
sigma models. Indeed, from a similar analysis one can conclude that    
there is no spontaneous CP symmetry breaking in \cp\ models at    
$\theta=0$.  This result can be checked for the \cpa\ model where the    
exact solution is known for $\theta=0$ \cite{zamm,pw}. 
    
The solution of the \cpa\ model at $\theta=0$ describes the    
interaction of a massive scalar particle in the adjoint representation    
of $SU(2)$, and CP symmetry is preserved confirming the result of the    
Vafa-Witten theorem.  
    
Another case where an exact solution is also known is the \cp\ model    
in the large $N$ limit \cite{adl}.  In this case the system also describes a    
weakly interacting and parity preserving massive scalar particle in    
the adjoint representation of $SU(N)$.  Let us analyse in some detail how    
the theorem works in this particular case. The effective action in a    
topological sector $q\neq 0$  with periodic boundary conditions,   
is given at leading order in $N$ by   
\begin{widetext}     
\be    
S_{\rm eff}(m,q)=-{N m^2 V\over 4 \pi}\log {4 \pi |q|\over \mu^2 V} -    
q N \log\displaystyle    
{\Gamma\left(\ha+{m^2 V\over 4 \pi |q|}\right)\over \sqrt{2\pi}} -    
{N m^2 V\over 2 g^2}    
\ee   
%
\end{widetext}     
\noindent   
where $2 m$ is the mass of the physical particles, $\mu=m \ {\rm exp}    
(2 \pi/g^2)$ is an energy scale fixing the value of the coupling    
constant $g$, $V=LT$ is the volume of the Euclidean space-time {\af T}$^2$ and 
$\Gamma$ is the Euler gamma function.  The effective action has a    
simple linear asymptotic regime for large values of the topological    
charge $q > m^2 V$,    
%
\be    
\displaystyle    
S_{\rm eff}(m,q)=-{N m^2 V\over 4 \pi} \log{4\pi |q| \over \mu^2 V}+    
\displaystyle{N |q|\over 2}\log 2 +    
\CO (1)    
\label{asymp}    
\ee    
%
\noindent   
The radius of convergence of the partition function $\CZ_\theta$    
in the pure imaginary axis $\theta=\dot\imath\vartheta$ is    
$\vartheta_c=N/2\log 2$. The critical value $\theta_c$ increases with    
$N$, and in the large $N$ limit the whole pure imaginary    
axis is free of Lee-Yang singularities in agreement with the general    
theorem.  In the complex $\theta$-plane the radius of convergence    
around the origin is larger than $\pi/4$ even in the infinite volume   
limit, which also confirms  the Vafa-Witten theorem    
in the large $N$ limit.  Now, the theorem holds not only in that limit,    
but for any finite value of $N$ as well.    
    
Finally, we remark that the asymptotic behavior of the effective    
action for small values of the topological charge $q<m^2 V$   
is quadratic instead of linear in $q$,   
%
\be    
\displaystyle    
S_{\rm eff}(m,q)={N m^2 V\over 4 \pi}    
(1-\log {m^2\over \mu^2})+ \displaystyle {N \pi  q^2\over 6m^2 V}    
+ \CO (q^4).    
\label{asymp2}    
\ee    
%
\noindent   
 Since $V=L T$ the change of    
asymptotic behavior from (\ref {asymp}) to (\ref {asymp2}) can be    
associated with a finite temperature $\beta=1/T$ crossover from the    
low temperature regime $\beta>m^2 L/|q|$ to the high temperature    
regime $\beta<m^2 L/|q|$ and cannot be related to any phase transition    
\cite{aff}. One might expect a similar behavior in QCD, although in    
that case there is a finite temperature phase transition \cite{kpt}.    

The arguments used above cannot exclude the existence of  higher-order 
phase transitions which do not  require spontaneous parity
symmetry breaking. Indeed, if $\rho_0=0$ but 
$\rho(z) \sim \rho_1\, z_2^\alpha \delta(z_1- m z_2^\beta)$
with $0<\alpha\leq 1$ for  $|z|\ll 1$, the vacuum energy density 
presents a second order singularity  and the topological
susceptibility diverges at $\vartheta=0$. This behavior is
compatible with the Vafa-Witten inequality and  might explain the 
behavior of the topological susceptibility of the \cpa\ model 
in  numerical simulations \cite{lh}. 

We thank J. Barb\'on,  I. Cirac, V. Laliena, E. Seiler and E. Witten for illuminating discussions. 

\end{document}